\documentclass[12pt,useAMS,useAASTEX,usenatbib]{aastex}
\usepackage{emulateapj5}

\makeatletter

\newenvironment{apjemufigure}{%
\def\@captype{figure}%
\noindent\begin{minipage}{0.999\linewidth}\begin{center}}
{\end{center}\end{minipage}}
\makeatother
\def\l{\ell}

\def\healpix{H{\sc ealpix }}
\def\glesp{G{\sc lesp }}

\def\wmap{\hbox{\sl WMAP~}}

\def\etal{et al.}
\newcommand{\bi}[1]{\mbox{\boldmath $#1$}}
\def\alm{a_{\l m}}
\def\Ylm{Y_{\l m}}
\def\Cl{C_{\l}}

\def\const{\rm const}

\def\summ{\sum_{m=-\l}^{\l}}
\def\suml{\sum_{\l=0}^{\infty}}
\def\lm{\l m}

\def\Cs{{\bf C}}
\def\Si{{\bf S}}

\def\R{{\bi R}}
\def\r{{\bi r}}

\def\S{{\bf S}}

\newcommand{\nbi}{{Niels Bohr Institute, Blegdamsvej 17,
DK-2100 Copenhagen, Denmark}}

\newcommand{\sao}{{Special Astrophysical Observatory, Nizhnij Arkhyz,
Karachaj-Cherkesia, 369167, Russia}}

\slugcomment{Submitted to {\it The Astrophysical Journal}}

\begin{document}

\title{Primordial magnetic field and non-Gaussianity of the 1-year
  {\it Wilkinson Microwave Anisotropy Probe} ({\it WMAP}) data}
 
\author{
Pavel D. Naselsky\altaffilmark{1},
Lung-Yih Chiang \altaffilmark{1},
Poul Olesen\altaffilmark{1},
Oleg V. Verkhodanov\altaffilmark{2}
}

\altaffiltext{1}{\nbi}
\altaffiltext{2}{\sao}

\email{naselsky@nbi.dk}


\begin{abstract}
Alfven turbulence caused by statistically isotropic and homogeneous
primordial magnetic field induces correlations in the cosmic microwave
background anisotropies. The correlations are specifically between
spherical harmonic modes $a_{\l-1,m}$ and $a_{\l+1,m}$. In this paper
we approach this issue from phase analysis of the CMB maps derived
from the \wmap data sets. Using circular statistics and return phase
mapping we examine phase correlation of $\Delta \l=2$ for the 
primordial non-Gaussianity caused by the Alfven turbulence at the epoch of
recombination. Our analyses show that such specific features from
the power-law Alfven turbulence do not contribute significantly in
the phases of the maps and could not be a source of primordial
non-Gaussianity of the CMB.  

\end{abstract}
\section{Introduction}
After the release of the 1-year \wmap data for analysis
\citep{wmap,wmapmap,wmapfg,wmapsystematics,wmappw}, one of the most exciting
 area of investigation is in the statistical characterization of the
 CMB signal. After the \wmap science team's report on the
 observational constraint on the quadratic non-Gaussianity of the
 anisotropy of the CMB \citep{wmapng}, \citet{tacng,coleskuiper,ndv3,ndv4,santenderng,park,eriksena,hansen,wandelt} have reported detections of non-Gaussianity by
various methods for the CMB maps derived from the \wmap
data. Unfortunately, the origin of such non-Gaussian features,
detected by different methods, is still unknown.
In \citet{tacng,ndv3,ndv4,eriksena}, they point out that the
non-Gaussian features might come from the foregrounds, whereas
\citet{hansen} argue in favor of systematic effects. Moreover,
different methods detect various properties of non-Gaussian features
from localized peculiar spots \citep{santenderng} to global
north-south asymmetry in the \wmap signal and COBE map as well
\citep{eriksenb}.

There is another point of view on the issue of the \wmap
non-Gaussianity. It comes from the theory of turbulence in
magnetohydrodynamics (MHD). Using the so-called extended
self-similarity (ESS) method, \citet{bs03,bs04}
point out that the statistical 
properties of angular increments $\delta T_r=T(\R+\r)-T(\R)$, where
$\r$ is the vector connecting two pixels $\R+\r$ and $\R$ of the map,
have the following relation in MHD $\langle|\delta T_\r|^p\rangle
\sim r^{{\xi}_p}$ for high order moments $p=4, 6 \ldots 12$. For
those moments, the ESS is clearly
detected both in the COBE and the \wmap data : $\langle |\delta
T_r|^p\rangle \sim  \langle \delta T_r^4\rangle^{{\xi}_p}$ where
${\xi}_p=\frac{p}{8} +1 -(\frac{1}{2})^{\frac{p}{4}}$. Moreover, in
\citet{beo,dky,mack,ssb,chen} it was shown that
primordial magnetic field, one of the most significant relics from
inflation, can generate
vorticity of Alfven waves before and
during the epoch of hydrogen recombination. It interacts with the CMB
photons, producing non-adiabatic tail of the CMB anisotropy and
polarization. Assuming statistical homogeneity and isotropy on
primordial magnetic field, it is shown that such Alfven turbulence
shall have non-Gaussian properties, because of quadratic dependence of
the vorticity amplitude on the magnetic strength $\bf{B}$. Namely,
such a magnetic field induces correlation between the $a_{\l-1,m}$ and
$a_{\l+1,m}$ multipole coefficients of the CMB temperature anisotropy
expansion by the spherical harmonics. 

An intriguing issue then follows: has \wmap observed the cosmological
Alfven turbulence? Recently this issue is discussed in
\citet{chen}, which exploits the
correlations between $a_{\l-1,m}$ and $a_{\l+1,m}$ of the CMB caused
by the vorticity of Alfven waves in order to put constraints on
the strength of the magnetic field: $|\bf{B}|< 15$ nG for the spectral
power index $n=-5$, and  $|\bf{B}|< 1.7$ nG for the spectral power
index $n=-7$. 

In this paper we discuss this issue from another aspect:
could the detected non-Gaussianity from the \wmap data be related to the Alfven
turbulence? The basic idea is that even if the Alfven turbulence is small
in amplitude in the CMB signal, it shall manifest itself  in the
\wmap maps between the phases $\phi_{\l-1,m}$ and
$\phi_{\l+1,m}$. Phase correlation between Fourier modes has been
investigated in relation to large-scale structure formation of the universe
\citep{scherrer,phaseentropy,mapping} and also applied as a test on
non-Gaussianity \citep{meanchisquare,tacng,coleskuiper} based on the
random phase hypothesis as a practical definition of Gaussian random fields
\citep{bbks,be}. \citet{phspointsource} use neighboring phase
correlations to extract extragalactic point sources.

Assuming that the CMB signal is composed of pure Gaussian signal and
sub-dominant vorticity tail, we use circular statistics on phases for
such a signal in order
to place constraint on the power spectrum of Alfven turbulence at each
multipole number $\l$. We will show that circular statistics allow us
to decrease contamination of the dominant (Gaussian) part of the
signal in order to investigate possible contamination from the
non-Gaussian sub-dominant part.

\section{Circular statistics of the Alfven turbulence}
For statistical characterization of temperature fluctuations on 
a sphere we express each signal (either CMB or foreground components)
as a sum over spherical harmonics: 
\begin{equation}
\Delta T(\theta,\varphi)=\suml \summ |\alm|e^{i\phi_{\l m}} \Ylm
(\theta,\varphi), 
\label{eq1}
\end{equation}
where $|\alm|$ and $\phi_{\l m}$ are the moduli and phases of the
coefficients of the expansion.

Homogeneous and isotropic CMB Gaussian random fields (GRFs), as a result
of the simplest inflation paradigm, possess Fourier modes whose real
and imaginary parts are Gaussian and mutually independent. The statistical
properties are then completely specified by its angular power spectrum
$\Cl^{cmb}$,  
\begin{equation}
\langle  a^{cmb}_{\l^{ } m^{ }} (a^{cmb}_{\l^{'}m^{'}})^{*}
\rangle = \Cl^{cmb} \; \delta_{\l^{ } \l^{'}} \delta_{m^{} m^{'}}.  
\label{eq2}
\end{equation}
In other words, from the Central Limit Theorem their phases
\begin{equation}
\Psi^{cmb}_{\l m}=\tan^{-1}\frac{\Im (\alm^{cmb})}{\Re (\alm^{cmb})}
\label{eq3}
\end{equation}
are randomly and uniformly distributed at the range
$[0,2\pi]$. We hereafter denote with {\it G}  as the
pure Gaussian tail of the CMB signal. For the combined signal 
\begin{equation}
\alm=G_{\l m}+V_{\l m},
\label{eq4}
\end{equation}
where $G_{\l m}$ is the Gaussian tail and $V_{\l m}$ is the vortex
tail. We can write down 
\begin{equation}
\alm=|\alm|\exp(i\Psi_{\l m}),
\label{eq4}
\end{equation}
where 
\begin{eqnarray}
|\alm|^2&=&|G_{\l m}|^2 + |V_{\l m}|^2 + 2|G_{\l m}||V_{\l
  m}|\cos(\Psi^G_{\l m}-\Psi^V_{\l m}); \nonumber\\
\tan\Psi_{\l m}&=&\frac{|G_{\l m}|\sin\Psi^G_{\l m} +|V_{\l
  m}|\sin\Psi^V_{\l m}}{|G_{\l m}|\cos\Psi^G_{\l m} +|V_{\l
  m}|\cos\Psi^V_{\l m}} 
\label{phas}
\end{eqnarray}
and   $\Psi^G_{\l m}$, $\Psi^V_{\l m}$ are the phases of Gaussian and
 vortex at the $(\l, m)$ harmonics, respectively. 

In \citet{dky} (hereafter DKY) it is shown
that for the CMB anisotropy produced by the Alfven turbulence, the
properties of the power spectrum and phases are different from the
adiabatic modes, which corresponds to the Gaussian tail of the
signal. Namely, for the power spectrum $\Cl(m)=\langle
\alm\alm^{*}\rangle$ and the auto-correlator $D_\l(m)=\langle
a^{}_{\l-1,m}a^{*}_{\l+1,m}\rangle$ from DKY we get
\begin{eqnarray}
\Cl(m)=&& A\frac{\Gamma(-n-1)\Gamma(\l+\frac{n}{2}+\frac{3}{2})}
{\Gamma(-\frac{n}{2})^2\Gamma(\l-\frac{n}{2}+\frac{1}{2})}\times \nonumber \\
&&\frac{2\l^4 +4\l^3 -\l^2 -3\l +(6 -2\l
  -2\l^2)m^2}{(2\l-1)(2\l+3)};\nonumber \\ 
D_\l(m)=&& \frac{2A}{|n+1|}\frac{(\Gamma(-n-1) \Gamma(\l+\frac{n}{2} +
  \frac{3}{2})}{\Gamma(-\frac{n}{2})^2 \Gamma(\l-\frac{n}{2} +
  \frac{1}{2})} (\l-1)(\l+2) \times  \nonumber \\ &&
\left[\frac{(\l+m+1)(\l-m+1) (\l+m) (\l-m)}{(2\l-1) (2\l+1)^2(2\l+3)}
\right]^{\frac{1}{2}}, 
\label{power}
\end{eqnarray}
where $A$ is the normalization constant and $-7 \le n \le -1$ is the
power spectral index of the Alfven turbulence. Below we discuss the
models with $n=-3$, $-5$ and $-7$, for which the mean power spectrum
over $m$ direction in Eq.(\ref{power}) are
$\overline{C}_\l$, $\overline{D}_\l \propto \l^{n+3}$. Note that
the model of $n=-5$ corresponds to the power spectrum
$\overline{\Cl} \propto \l^{-2}$, which at the multipole range of
$5<\l< 20$ has the same feature as that of the CMB for adiabatic
perturbation. For the model of $n=-7$, on the other hand, the power of
vertex tail is $\overline{\Cl} \propto \l^{-4}$ and increase rapidly
if $\l\rightarrow 0$, which is typical for the diffuse extragalactic
foregrounds. For $n=-3$ the vertex power $\overline{\Cl}\sim \const$ mimic 
the power of extragalactic point source component. Therefore, simple
statistics based on the estimation of the $\overline{D}_\l$-th moment
from the \wmap data is potentially misleading, because of possible
contribution from such kind of foregrounds as sub-dominant components to
the signal.

The method we propose in this paper is based on the statistical
characterization: non-Gaussianity
of the vortex part of the CMB signal and related with correlation of
the phases of $(\l-1,m), (\l+1,m)$
harmonics. The correlation vanishes for pure Gaussian CMB
signal, whereas it is significant for the vortex component. We
therefore, introduce some specific functions of phases which minimize
the contribution from non-correlated Gaussian signal and
maximize the non-Gaussian tail of the phases. For this purpose we use
trigonometric moment statistics to counter the circular nature of
phases (\citet{fisher}, see also \citet{ndv3}). Let us define the following trigonometric moments:
\begin{eqnarray}
\Cs(\l)&=&\frac{1}{\l-1}\sum_{m=1}^{\l-1}
\cos(\Psi_{\l-1,m}-\Psi_{\l+1,m}); \nonumber\\
\Si(\l)&=&\frac{1}{\l-1}\sum_{m=1}^{\l-1}
\sin(\Psi_{\l-1,m}-\Psi_{\l+1,m}); \nonumber\\
r^2(\l)&=&{\Cs^2(\l)} +{\Si^2(\l)};\nonumber\\
 R&=&\frac{1}{\l_{\max} -\l_{\min}+1}\sum_{\l_{\min}}^{\l_{\max}}r(\l),
\label{def2}
\end{eqnarray}
The reason for such statistics is clear. Simple algebra leads to the
following properties of the phases:
\begin{eqnarray}
\cos(\Psi_{\l-1,m}-\Psi_{\l+1,m})&=&\nonumber\\
|g_{{\l-1,m}}||g_{{\l+1,m}}| && \cos(\Psi^G_{\l-1,m}-\Psi^G_{\l+1,m})
\nonumber\\ 
+|v_{{\l-1,m}}||v_{{\l+1,m}}| &&\cos(\Psi^V_{\l-1,m}-\Psi^V_{\l+1,m})
\nonumber\\ 
+|g_{{\l-1,m}}||v_{{\l+1,m}}| &&\cos(\Psi^G_{\l-1,m}-\Psi^V_{\l+1,m})
\nonumber\\ 
+|g_{{\l+1,m}}||v_{{\l-1,m}}| &&\cos(\Psi^G_{\l+1,m}-\Psi^V_{\l-1,m})
\label{def3}
\end{eqnarray}
where $|g_{\l m}|=|G_{\l m}|/|\alm|$,$|v_{\l m}|=|V_{\l
m}|/|\alm|$. As one  can see from Eq.(\ref{def3}) the first
term is proportional to $D_\l(m)=\langle
a^{}_{\l-1,m}a^{*}_{\l+1,m}\rangle$ moment for the
pure Gaussian tail of the signal, which should be vanish, the second
term is $D_\l(m)$ for the vortex, and last 
two terms correspond to correlations between $G$ and $V$, which should be
statistically negligible after summation over $m$.
If the vortex component is sub-dominant, then in Eq.(\ref{phas}) and
Eq.(\ref{def3}) $|V_{\l m}|\ll|G_{\l m}|$ for all $\l,m$ and $|\alm|$
is given by Eq.(\ref{phas}). This means that $|g_{\l m}|\simeq 1$ and
$|v_{\l m}|\ll 1$. However, because of finite number of $m$ modes and  
especially for low $\l$ range, correlations can manifest
themselves between $(\l-1,m)$, $(\l+1,m)$ harmonics. In order of magnitude
this effect can be estimated by the following way. For pure Gaussian
signal with non-correlated phases $\phi_{\l m}$ one obtains 
\begin{equation}
\frac{1}{\l}\sum_{m=1}^{\l}\cos\phi_{\l m}\simeq
\frac{1}{\sqrt{\l}};
\hspace{0.5cm}\frac{1}{\l}\sum_{m=1}^{\l}\sin\phi_{\l m}\simeq
\frac{1}{\sqrt{\l}}. 
\label{rand} 
\end{equation}
Then,for the moduli $|\alm|$ from Eq.(\ref{phas}) one gets
\begin{eqnarray}
|\alm|^2 & \simeq & |G_{\l m}|^2 +|V_{\l m}|^2 \nonumber \\
 {\rm if}& & ||V_{\l m}|\gg |G_{\l m}|/{\sqrt{\l}}; \nonumber\\
|\alm|^2 & \simeq & |G_{\l m}|^2 +2|G_{\l m}||V_{\l m}|\cos(\Psi^G_{\l
 m}-\Psi^V_{\l m}), \nonumber\\
{\rm if} && |V_{\l m}|\ll |G_{\l m}|/{\sqrt{\l}}.
\label{modulo} 
\end{eqnarray}
For the asymptotic $|V_{\l m}|\gg |G_{\l m}|/{\sqrt{\l}}$,
\begin{eqnarray}
\Cs(\l) \sim \frac{1}{\sqrt{\l}}
-\frac{1}{\l^{\frac{3}{2}}}\sum_m\frac{|V_{{\l-1,m}}||V_{{\l+1,m}}|}
{|G_{{\ell-1,m}}| |G_{{\l+1,m}}|} + \nonumber\\
\frac{1}{\l}\sum_m\frac{|V_{{\l-1,m}}| |V_{{\l+1,m}}|}{|G_{{\l-1,m}}|
  |G_{{\l+1,m}}|} \cos(\Psi^V_{\l-1,m}-\Psi^V_{\l+1,m}).
\label{sincos} 
\end{eqnarray}
For the correlated phases of the vortex component, the last term in
Eq.(\ref{sincos}) is larger than the second and, if
$\Psi^V_{\l-1,m} \simeq \Psi^V_{\l+1,m}$, then 
\begin{equation}
\Cs(\l) \sim \frac{1}{\sqrt{\l}}+\frac{{\overline D}_\l}{{\overline G}^2_\l},
\label{cos}
\end{equation}
where ${\overline G}^2_\l$ is the power spectrum of the Gaussian tail.

For highly correlated vortex perturbations, therefore, the contribution to the
$\Cs$ function is significant if  $\frac{{\overline
    D}_\l}{{\overline G}^2_\l} \sqrt{\l}\ge 1$. At $\l =
100$,  $\frac{{\overline D}_\l}{{\overline G}^2_\l}\ge 0.1$, which
is exactly the DKY criterion for estimation of the magnetic field
amplitude for different values of the spectral index $n$. However,
if the observable data sets covering the range of multipoles up to
the Silk damping scale $\l_d$ for the vortex perturbations, $\l_d \sim
500$ \citep{dky}, then non-Gaussian features could be detectable for $ \frac{{\overline
  D}_\l}{{\overline G}^2_\l} \ge 0.045$. Moreover, one interesting feature
comes from power spectral index dependence of the vortex perturbations
mentioned above. For $n=-7$, ${\overline D}_\l\propto \l^{-4}$, while
${\overline G}^2_\l$ has the same asymptotic $\l^{-2}$ as the
$\Lambda$CDM \wmap best fit model at $5<\l<30$. For that model of the
power index, the most stringent constraint should come from statistics
of the low
multipoles from the \wmap data, then from high multipole
range $\l \sim \l_d$ and simple estimator could be $6 \div 7$ times bigger
than in DKY: $\frac{{\overline D}_\l}{{\overline G}^2_\l} \sim 0.6\div0.7$.

We would like to point out that the above-mentioned properties of the
$\Cs$ statistics provide natural explanations in terms of the
cosmic variance limit of error bars for any CMB
experiments. Assuming no systematic effect present in the data, for
low multipoles $\l < 100$ the cosmic variance limit corresponds to
$\delta \Cl/\Cl\simeq (f_{\rm sky} \l)^{-\frac{1}{2}}$, where $f_{\rm
  sky}$ is the sky coverage of the observation. The whole sky \wmap
coverage gives $f_{\rm sky}=1$. If some part of the CMB power spectrum
is related to vortex perturbations, then \citep{chen}
\begin{equation}
\Cl={\overline G}^2_\l + {\overline C}_\l\sim {\overline G}^2_\l+
|n+1|\left[\frac{\Gamma
    (-\frac{n+1}{2})}{\Gamma(-\frac{n}{2})}\right]^2 {\overline D}_\l.
\label{cosvar}
\end{equation}
Using the definition $\delta \Cl/\Cl=(\Cl-{\overline G}^2_\l)/{\overline
  G}^2_\l$, which corresponds to the contribution of the vortex perturbations
to the power spectrum, we have the following threshold of detectability
\begin{equation}
|n+1| \left[\frac{\Gamma(-\frac{n+1} {2})}{
 \Gamma(-\frac{n}{2})}\right]^2 {\overline D}_\l/{\overline G}^2_\l > \frac{
1}{\sqrt{\l}},
\label{cosvar1}
\end{equation}
which is in perfect agreement with our estimation Eq.(\ref{cos}).

Let us briefly discuss the properties of the $\Si$ statistics. Similar
to Eq.(\ref{def3}) one can prove that 
$\sin(\Psi_{\l-1,m}-\Psi_{\l+1,m})$ is given by Eq.(\ref{def3}) with
transition from cosine to sine function. However, because of $\sin$
dependency, if phases of the non-Gaussian tail are highly correlated, we
have degradation of the 
$|v_{{\l-1,m}}||v_{{\l+1,m}}|\sin(\Phi^V_{\l-1,m}-\Phi^V_{\l+1,m})$
term, even if $|V_{\l m}| \gg |G_{\l m}|/{\sqrt{\l}}$. So, if the
phases of the vortex perturbations $\Psi^V_{\l-1,m}-\Psi^V_{\l+1,m})$
are highly correlated, for the $\Si$ statistics we get
$\Si \sim 1/{\sqrt{\l}}$. Thus, the presence of vertex perturbation (as
all non-Gaussian signals) in the CMB data manifest itself as asymmetry
of the $\Cs$-$\Si$ statistics, while for the pure Gaussian signals it should be
symmetrical one. In order of magnitude this asymmetry for each mode $\l$ is
\begin{equation}
A(\l)=\frac{\Si(\l)}{\Cs(\l)}=\frac{1}{1+{\sqrt{\l}}\frac{{\overline
      D}_\l}{{\overline G}^2_\l}} \ll 0.5,
\label{asym}
\end{equation}
for highly correlated phases of vortex perturbations.
Addition to Eq.(\ref{asym}) we can define global asymmetry of the signal
\begin{equation}
A_g=\left|\frac{\sum_\l\Si(\l)}{\sum_\l\Cs(\l)}\right|.
\label{gasym}
\end{equation}
In terms of circular statistical variables \citep{fisher}, this global
asymmetry corresponds to the mean angle $\Theta=\tan^{-1} A_g$ for
orientation of all harmonics $(\l,m)$ in the CMB map. Needles to say,
for vertex perturbations with $n=-5$ global asymmetry is expected to
be in order of magnitude  
\begin{equation}
A_g=\frac{1}{1+\frac{{\overline D}_\l}{{\overline G}^2_\l}\l_{\max}},
\label{gasym1} 
\end{equation}
and can be detectable, if $\frac{{\overline D}_\l}{{\overline
    G}^2_\l}\l_{\max} \gg 1$.
At the end of this section we would like to mention one peculiar
asymptotic of the $A(\l)$ and  $A_g$ parameters
related with correlations of phases. From Eq.(\ref{sincos}) one
can find that correlation
$\Psi^V_{\l-1,m}-\Psi^V_{\l+1,m}=\pi/2, 3\pi/2$ leads to $A(\l),A_g >
    1$. So any peaks of the function
$A(\l),A_g\gg 1$ are the mark of $\pi/2, 3\pi/2$ correlations for
    the non-Gaussian tail of the CMB signal.
 Below in addition to definition Eq.(\ref{def2}) we introduce
the phase difference
\begin{equation}
D_{\l m}(\Delta \l)=\Psi_{\l-\frac{\Delta \l}{2},m}
-\Psi_{\l+\frac{\Delta \l}{2},m} 
\label{dif} 
\end{equation}
for even $\Delta \l$ and
\begin{equation}
D_{\l m}(\Delta \l)=\Psi_{\l+\Delta \l,m} -\Psi_{\l-\Delta \l-1,m}
\label{dif1} 
\end{equation}
for odd $\Delta \l$ and corresponding circular moments $\Si_{\Delta
  \l}(\l)$ and $\Cs_{\Delta \l}(\l)$ similar to Eq.(\ref{def2}). These
variables allow us to describe properties of the CMB phases for any $
  \Delta \l$ separations of the modes. 

\section{Circular statistics of the ILC, FCM and WFCM maps derived
 from the \wmap data}
In this section we apply the $\Si$, $\Cs$ statistics for the
three high resolution 
whole-sky maps: the Internal Linear Combination
Map (ILC) by the \wmap science team, the TOH foreground-cleaned map
(hereafter FCM) and the
TOH Wiener-filtered map (hereafter WFCM). The ILC and the FCM are
obtained from a weighted combination of the \wmap 5 maps at frequency
23, 33, 41, 61 and 94 GHz in order to separate the microwave
foreground. The WFCM is the map after foreground residue cleaning by
Wiener filtering. All the maps formally claim
resolution up to $\l=1024$, but none of them is correct for
investigation of the CMB properties. The ILC map, as is mentioned
at the \wmap website, is applicable for the foregrounds property
investigation at the range of multipoles $\l<300$. The FCM and WFCM
are smoothed  for the multipole range $\l>300$ by a Gaussian window
function and the CMB
signal is completely erased by the smoothing. However,
looking at the V and Q bands of the \wmap data and combining all the
foregrounds (synchrotron, free-free, dust emission) and the point
sources from the \wmap catalog, one can find that this map reproduces
the CMB signal outside the Kp2 Galactic cut mask extremely well at the
range of multipoles $\l \le 50$. 

The FCM and WFCM have well-defined scale of smoothing in the image
domain. However, smoothing does not change at all the phases of
the signal even down to the limit $\l=1024$. So, using FCM and WFCM
we can examine the phases at the whole range of
multipoles. One additional reason for using the FCM and WFCM maps is that
they are tested by ESS method, from which the
manifestation of the Alfven turbulence properties could be clearly
seen \citep{bs03,bs04}.

In Fig.\ref{deltal2} we plot for the
three maps (ILC, FCM and WFCM) the $\Si$-$\Cs$ phase diagrams for $\Delta
\l=(\l+1)-(\l-1)=2$. For comparison we include the
statistics of Gaussian random signal, which reproduce our expectation
of the phase properties. From the second row in Fig.\ref{deltal2} one
can see that ILC phases perfectly reproduce the properties of Gaussian signal
phases in terms of $\Si(\l)$, $\Cs(\l)$ and $A(\l)$ variables. One can
see that on the third and the fourth row, which corresponds to the FCM and
WFCM maps, starting from $\l \sim 100$ the 
$\Si$-$\Cs$ symmetry is broken, and while the $\Si$ follows
Gaussian statistics, its $\Cs$ and $A(\l)$ appear non-Gaussian. 

According to Fig.\ref{deltal2}, the phases of the FCM and
WFCM are in agreement with the expectation, described in previous
Section. We have significant increase of the $\Cs$ statistics, while
$\Si$ statistics are nearly the same as for the Gaussian
signal. However, next three Figures display clearly
non-Gaussianity. Fig.\ref{deltal1} are the circular statistics $\Si(\l)$,
$\Cs(\l)$ and $A(\l)$ for $\Delta \l=1$, and Fig.\ref{deltal4} and
Fig.\ref{deltal3} are for 
$\Delta \l=3$ and $\Delta \l=4$ respectively. According to DKY, for
all $\Delta \l \neq 2$ the phases of the signal
should not display any correlated features. One can also find 
remarkable symmetry in the FCM and WFCM between even and odd $\Delta
\l$ statistics (e.g. the $\Cs$ in Fig.\ref{deltal2} and
Fig.\ref{deltal4} and the $\Si$ in Fig.\ref{deltal1} and Fig.\ref{deltal3}).

\begin{apjemufigure}
\hbox{\hspace*{0.01cm}
\centerline{\includegraphics[width=1.\linewidth]{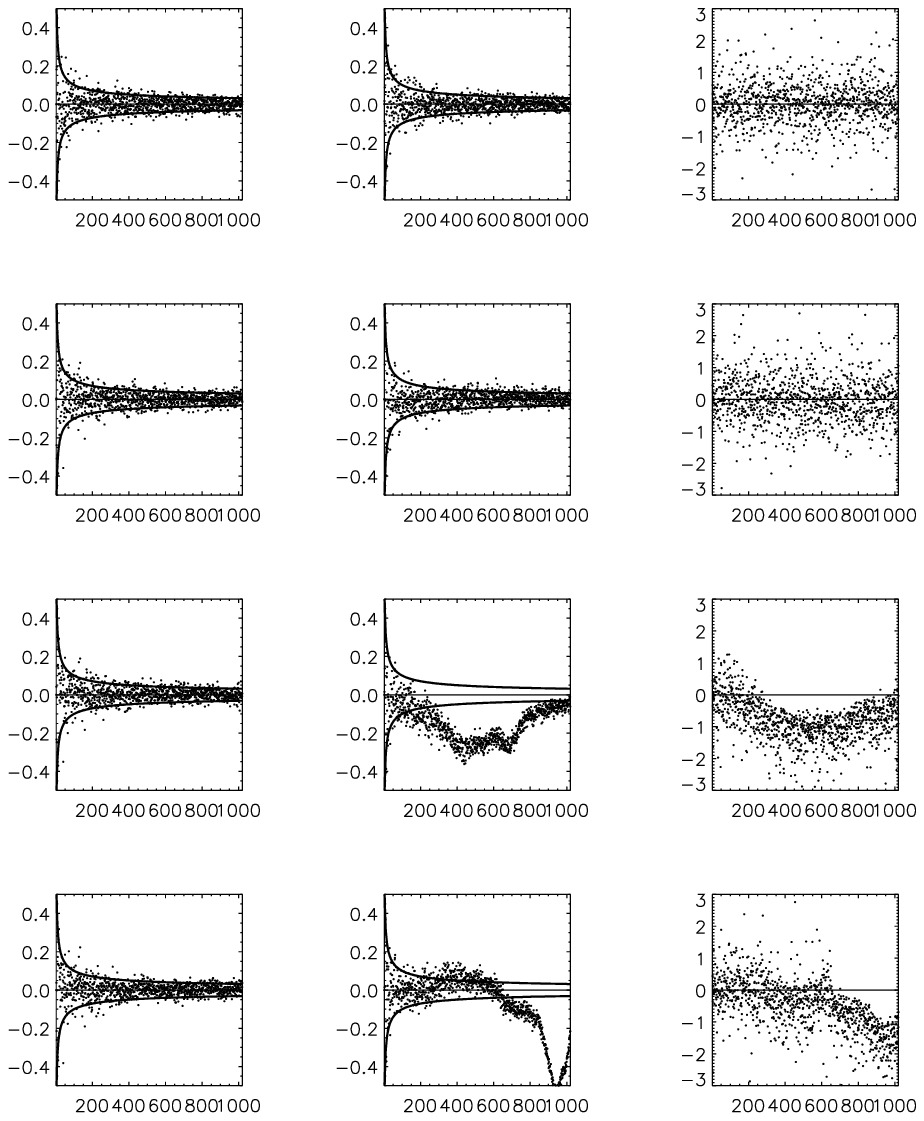}}}
\caption{Circular statistics on phase correlations with
$\Delta\l=2$. The columns (from left to the right) correspond to
$\Si(\l)$, $\Cs(\l)$ and $\log A(\l)$, respectively. The first row
is for a Gaussian realization (random phases). The second
row is the ILC map, the third the FCM map and the last the WFCM map.}
\label{deltal2} 
\end{apjemufigure}

\begin{eqnarray} 
sign\Cs_{\Delta \l} = - sign \, \Cs_2 \cos (\frac{\pi}{2}\Delta \l ),
{\rm for \,\, even\,\,}
\Delta \l \nonumber\\
sign\Si_{\Delta \l} = sign \, \Si_1\sin(\frac{\pi}{2}  \Delta \l),
{\rm for \,\, odd\,\,} \Delta \l.
\label{delt}
\end{eqnarray}
This symmetry reflects the symmetry of the phases $D_{\Delta\l}(\l)$, namely

\begin{equation}
\Psi_{\l+\frac{\Delta \l}{2},m}-\Psi_{\l-\frac{\Delta \l}{2},m}=\frac{\pi}{2}\Delta \l +\pi
\label{phcos}
\end{equation}
for even $\Delta \l$ and
\begin{equation}
\Psi_{\l+\Delta \l,m}-\Psi_{\l-\Delta \l-1,m}=\frac{\pi}{2}\Delta \l 
\label{phsin}
\end{equation}
for odd $\Delta \l$ and for all values  of $m$.

\section{Non-linear statistics}
In this section in order to investigate the properties of
non-Gaussianity, we apply non-linear statistics $r^2(\l)$ for the ILC,
FCM and WFCM phases. From Eq.(\ref{def2}) one obtains
\begin{equation}
r^2(\l)=\frac{1}{\l-1}+ \sum_m\sum_{n\neq m} \frac{\cos[D_{\l m}
  (\Delta \l)- D_{\l n}(\Delta \l)]}{(\l - 1)^2}
\label{r}
\end{equation}
That means, that $r$-statistics are sensitive to $m$ dependence of the
phase difference for given
values $\l$ and $\Delta \l$.

\begin{apjemufigure}
\hbox{\hspace*{-0.01cm}
\centerline{\includegraphics[width=1.\linewidth]{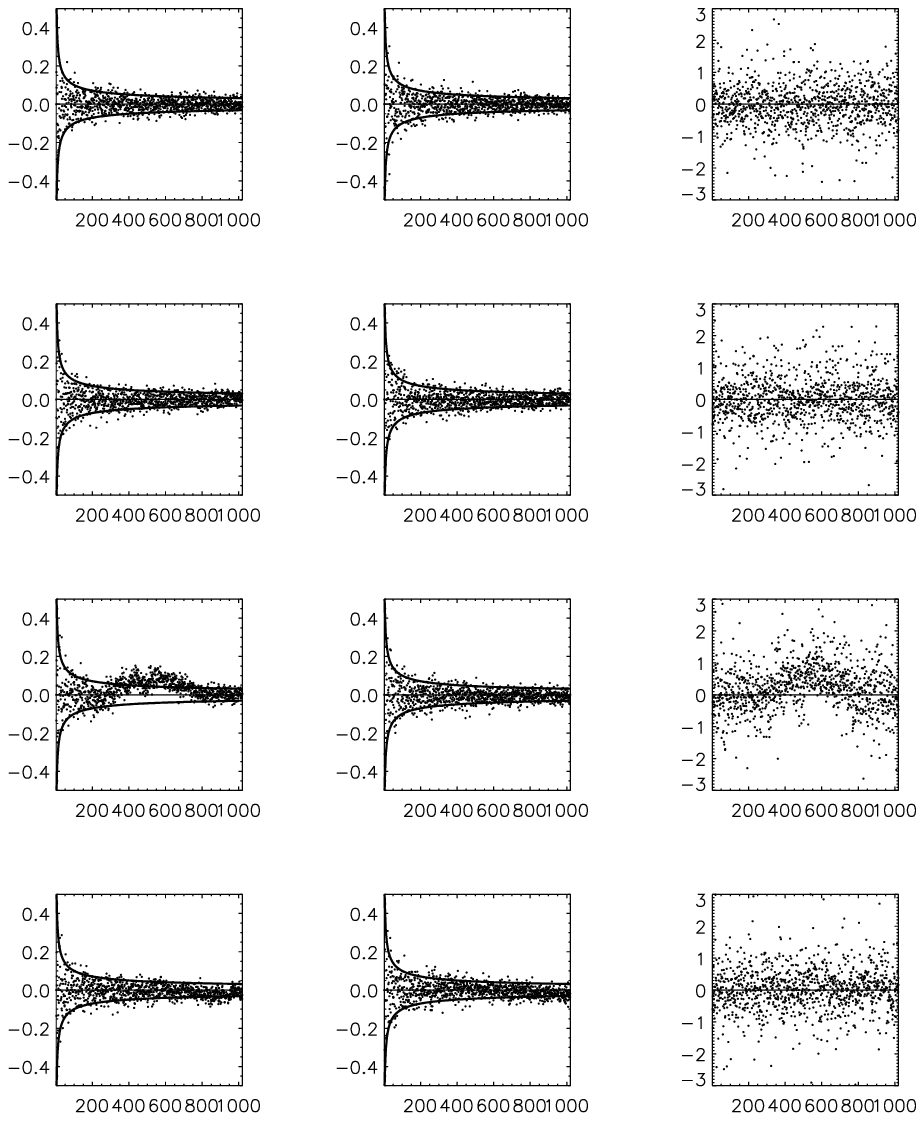}}}
\caption{The same as in Fig.1, but for $\Delta\l=1$.}
\label{deltal1} 
\end{apjemufigure}

\begin{apjemufigure}
\hbox{\hspace*{-0.01cm}
\centerline{\includegraphics[width=1.\linewidth]{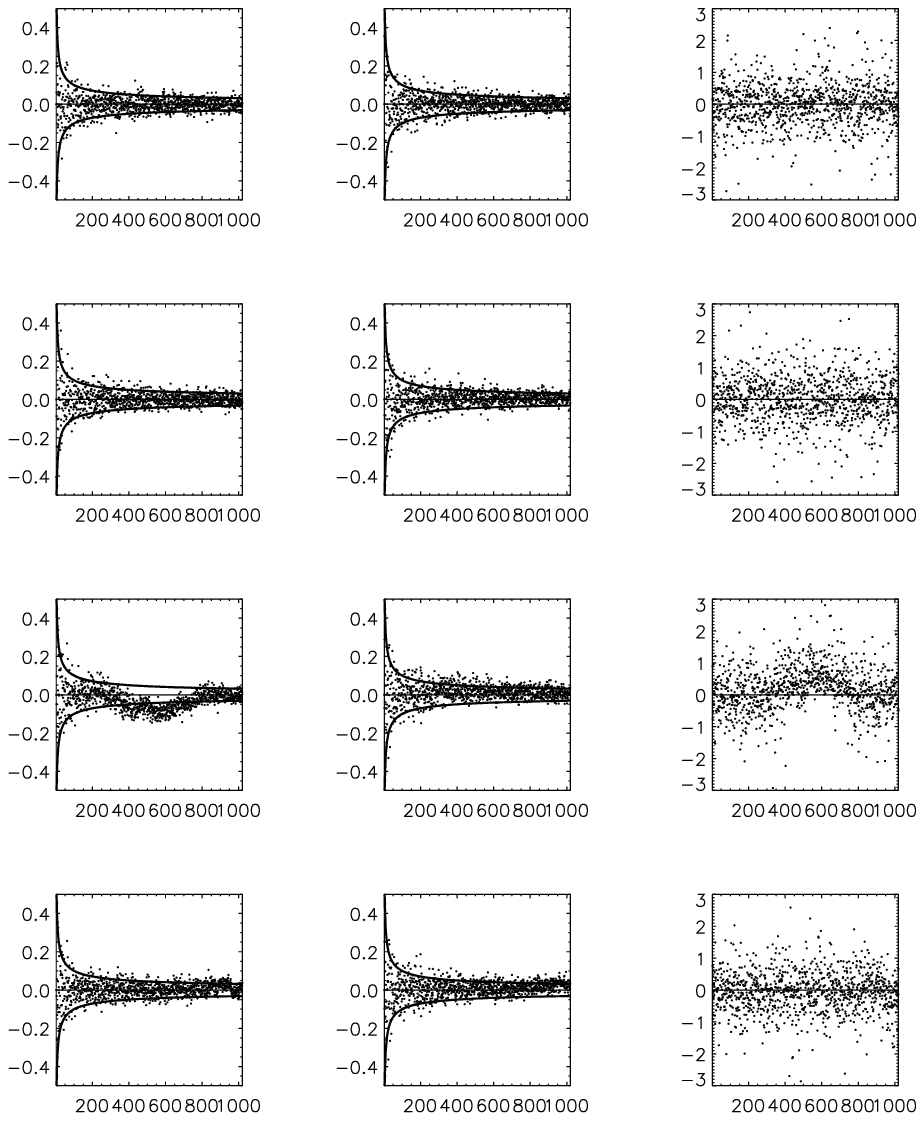}}}
\caption{The same as in Fig.1, but for $\Delta\l=3$.}
\label{deltal3} 
\end{apjemufigure}

\begin{apjemufigure}
\hbox{\hspace*{-0.01cm}
\centerline{\includegraphics[width=1.\linewidth]{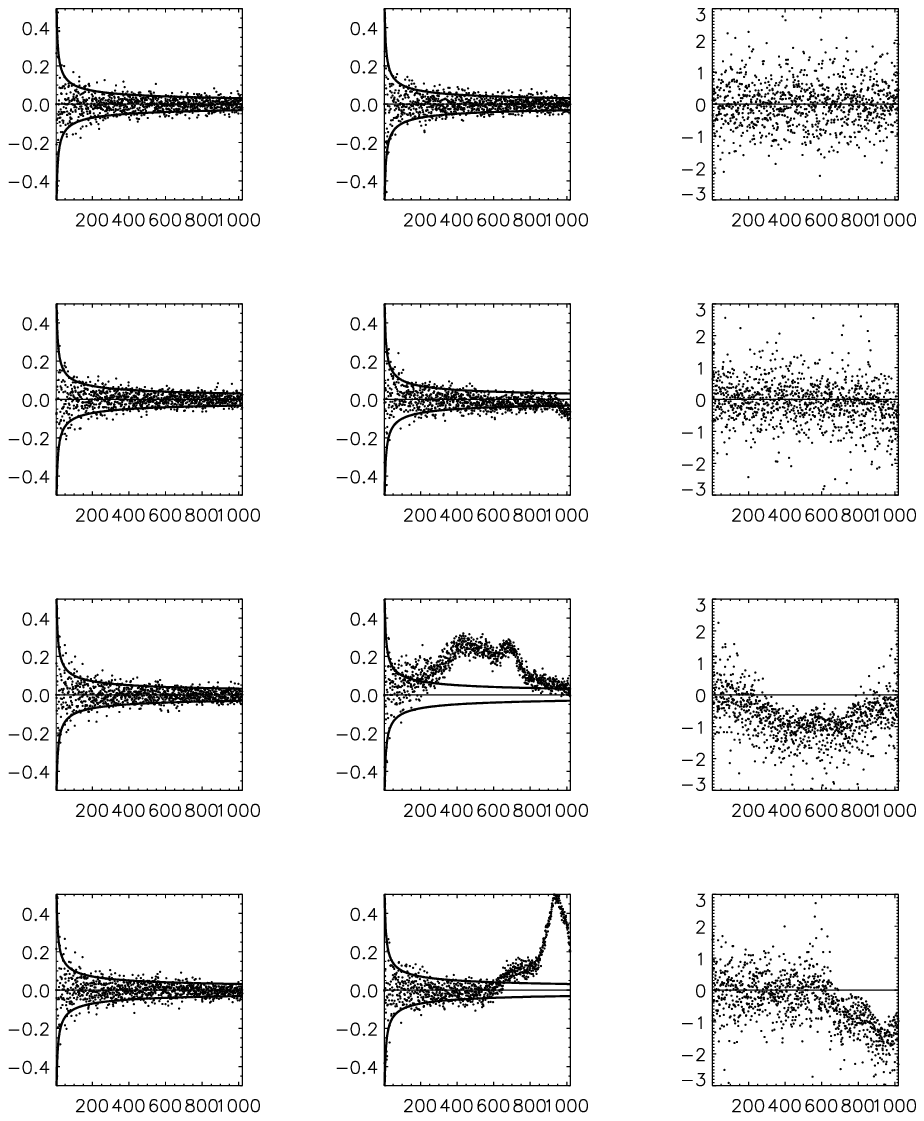}}}
\caption{The same as Fig.1, but for $\Delta\l=4$.}
\label{deltal4} 
\end{apjemufigure}

This statistic has well-defined properties. 
If the signal is highly correlated and $D_{\l m}(\Delta
\l)\rightarrow 0$ then $r^2(\l)\rightarrow 1$. If the phases are
non-correlated, however, $r^2(\l)\rightarrow \l^{-1}$. In
Fig.\ref{sim5} and \ref{sim6} we plot $r^2(\l)$ versus $\l$ for the same model
and the same order as in Fig.\ref{deltal2}. For all even $\Delta \l$ the
corresponding $r^2(\l)$ are similar to Fig.\ref{sim5}
and for all odd $\Delta \l$ the $r^2$ statistics follow Fig.\ref{sim6}.

Like $\Si$ and $\Cs$ statistics, $r^2(\l)$ statistics reflect
non-Gaussian features for $\Delta \l=1, 2, 3 \ldots$, which
allow us to conclude that non-Gaussianity of FCM and WFCM does not
relate with the vortex turbulence at the
epoch of recombination. Moreover, in the next Section we will show,
how the symmetry of the $\Si$ and $\Cs$ statistics
allow us to detect the source of non-Gaussianity of FCM and WFCM.

\section{Local defects of the ILC, FCM and WFCM maps}
In order to detect the nature of the non-Gaussianity of the maps with
different powers of signals, we will use the power filter
$P(\l)=1/|\alm|$, proposed by \citet{gorski} and
\citet{powerfiltration}. This linear
filter transforms any maps with $\alm$ coefficients of spherical
harmonic expansions to maps with
the same amplitude $(=1)$ at each mode $\l,m$, but preserving all the
phases $\Psi_{\l m}$ and their correlations.
\begin{equation}
N_{\l m}=P(\l)\alm=\exp(i\Psi_{\l m})
\label{pow}
\end{equation}

\begin{apjemufigure}
\hbox{\hspace*{-0.01cm}
\centerline{\includegraphics[width=1.\linewidth]{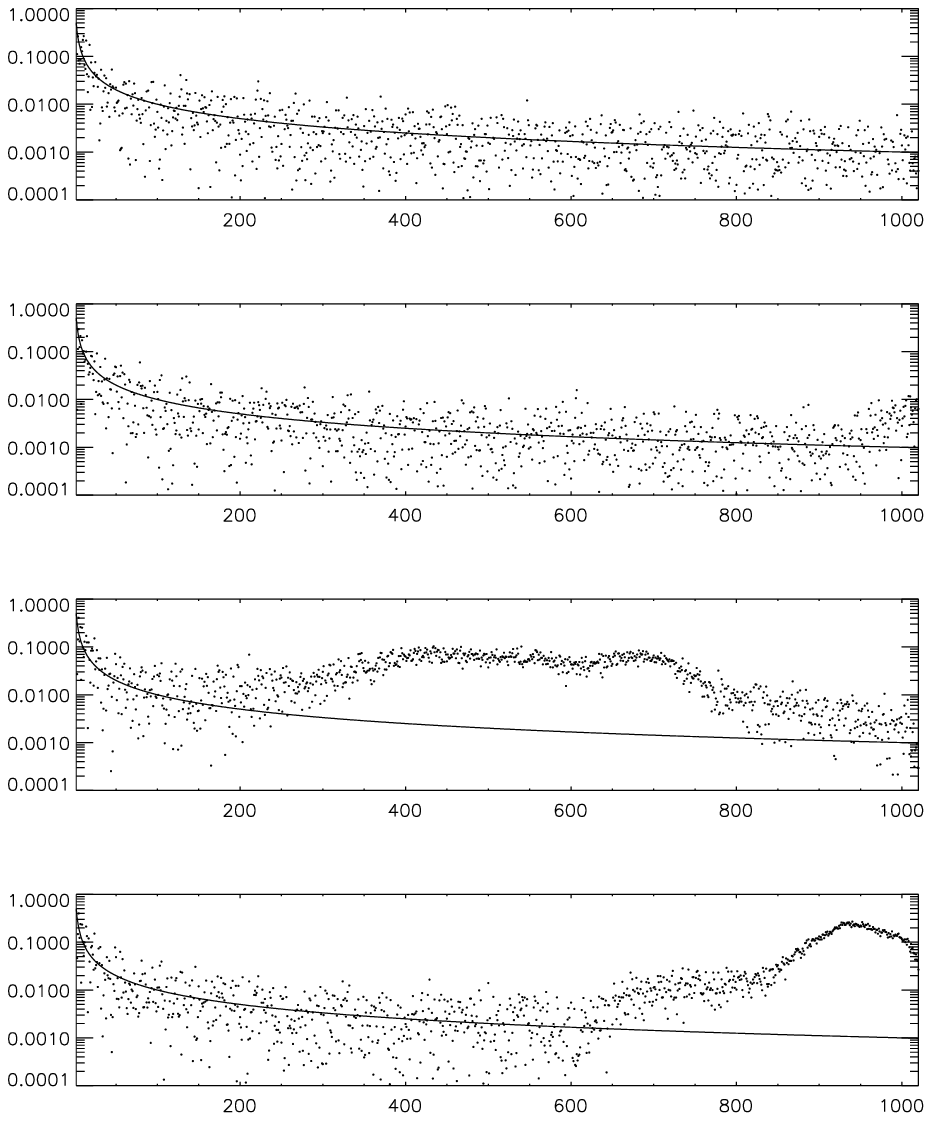}}}
\caption{ $r^2(\l)$ statistics for  $\Delta\l=2$ for (from top to
bottom) a Gaussian random realization, ILC, FCM and WFCM,
respectively. Solid line corresponds to $r^2(\l)=\l^{-1}$ asymptotic.}
\label{sim5} 
\end{apjemufigure}

\begin{apjemufigure}
\hbox{\hspace*{-0.01cm}
\centerline{\includegraphics[width=1.\linewidth]{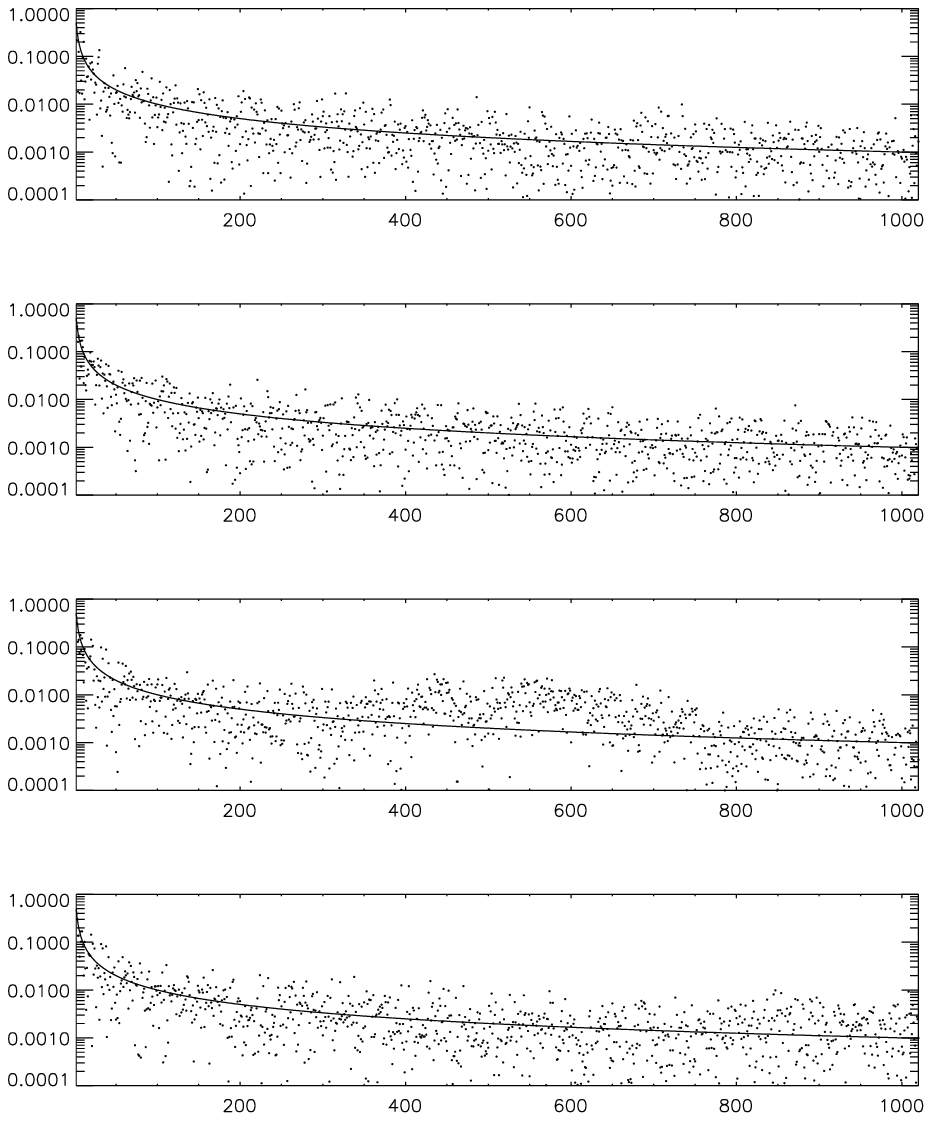}}}
\caption{The same as in Fig.\ref{sim5}, but for $\Delta\l=1$.}
\label{sim6} 
\end{apjemufigure}

In Fig.\ref{whitened} we plot the power filtered ILC, FCM and WFCM maps
after power filtration for the range of multipoles up to $\l=1024$ in
order to show how power filter allows us to detect some of the local defects
of the maps close to the Galactic plane. In particular, one can easily
see that there are defects in the ILC map which was originally 
obscured by the amplitudes $|\alm|$. The clear cut features is
obviously not of astrophysical origin. One crude guess for such peculiar
shape in the middle of the ILC map is that the \wmap science team
might have used a direct mask at the center of the map where there are
pronounced contaminations from foregrounds and then used Wiener filtering,
combining with the best-fit power spectrum, to recover the signal in
that area. However, the morphology is related to phases \citep{morphology} and
such mask shall alter the morphology, which manifests itself in the
phases. On the other hand, one can see foreground residues in both the
FCM and WFCM maps. Qualitatively speaking, these are clear signatures of
non-Gaussianity. 
\begin{apjemufigure}
\hbox{\hspace*{-0.5cm}
\centerline{\includegraphics[width=0.92\linewidth]{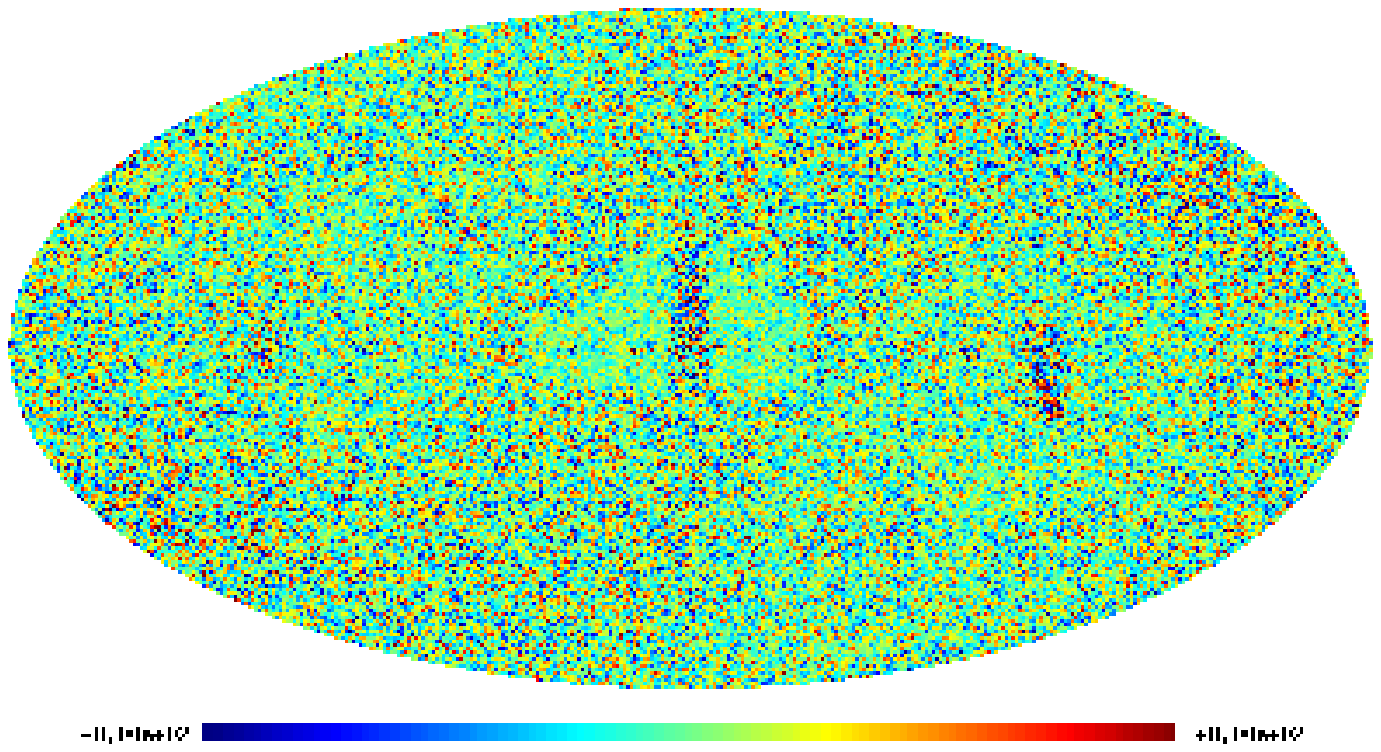}}}
\hbox{\hspace*{-0.5cm}
\centerline{\includegraphics[width=0.92\linewidth]{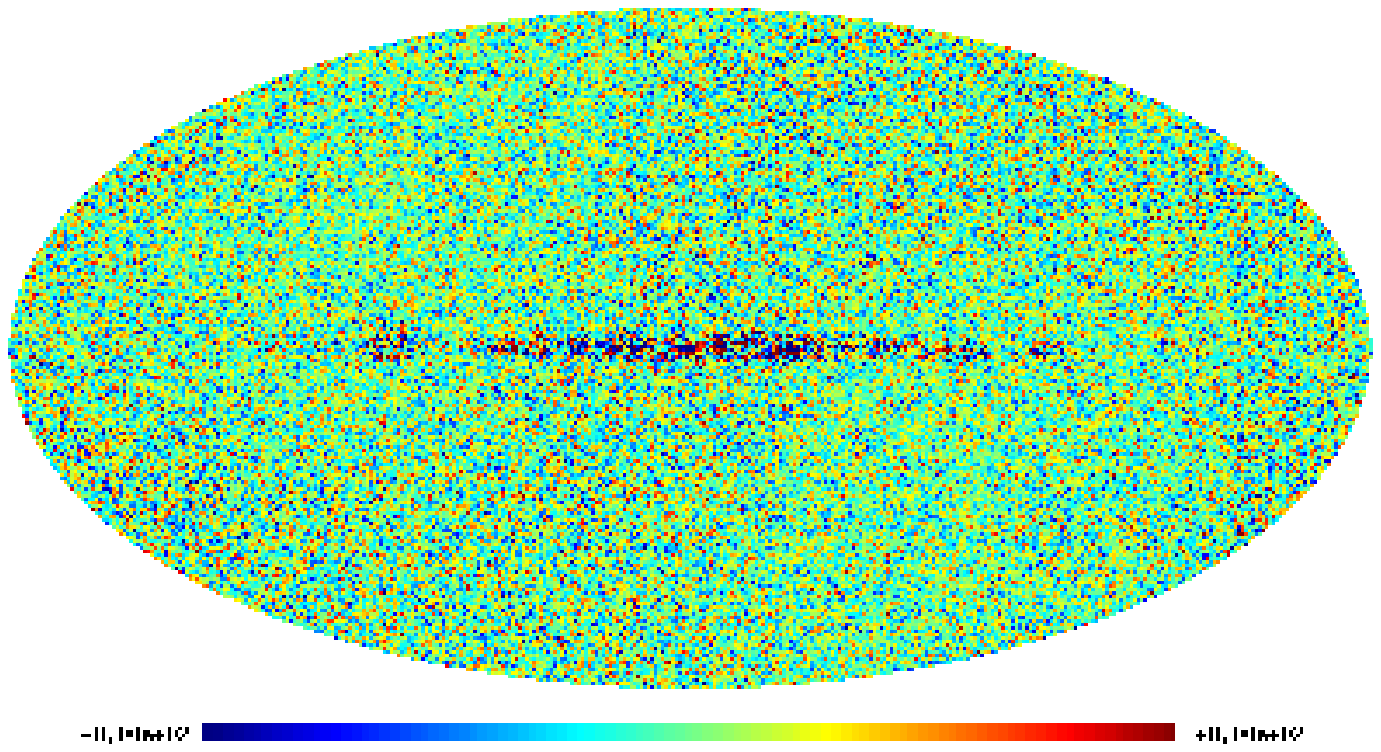}}}
\hbox{\hspace*{-0.5cm}
\centerline{\includegraphics[width=0.92\linewidth]{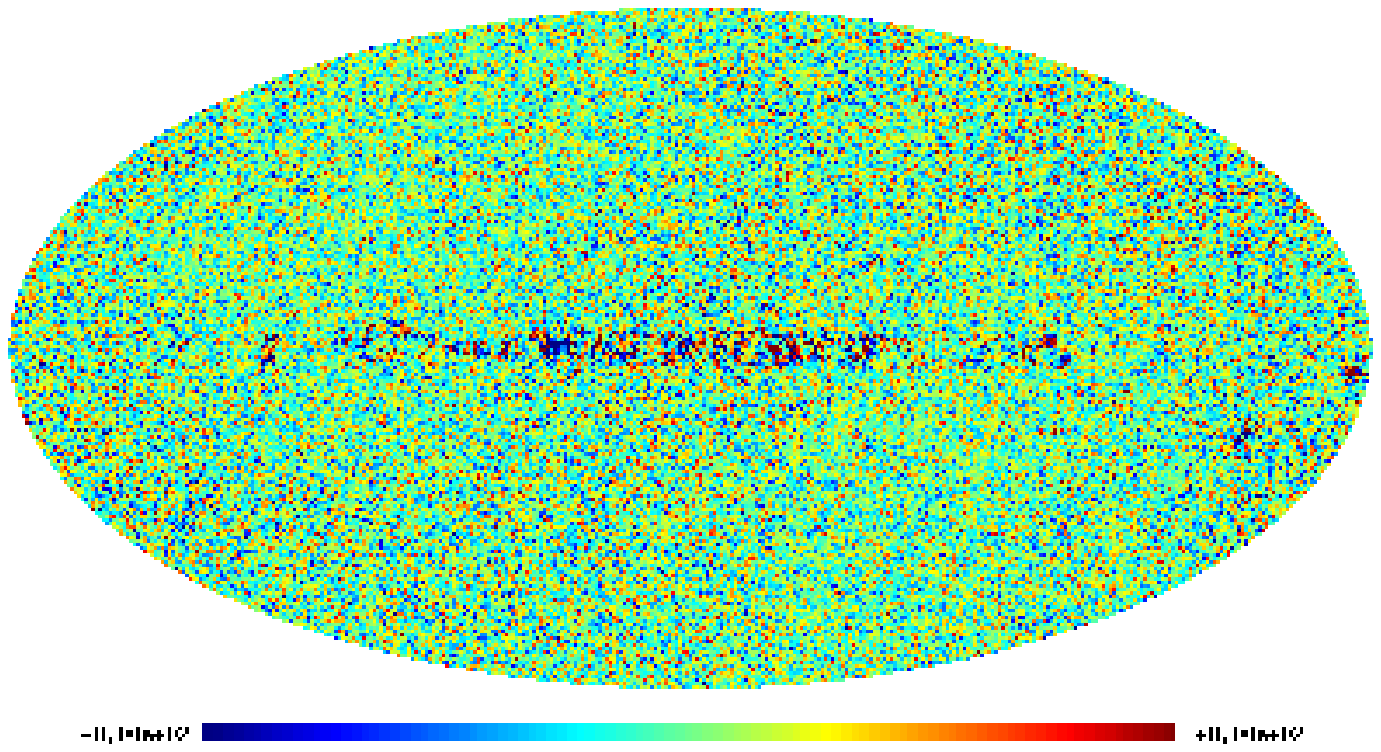}}}
\caption{The ILC (top), FCM (middle) and WFCM (bottom) power filtered
maps. These whiten images are to display how foreground cleaning and
the residues manifest themselves in phases, which are otherwise 
obscured by the amplitudes $|\alm|$. One can see the clear cut feature
in the middle of the ILC map and galactic and point source
contaminations for FCM and WFCM. Qualitatively, these are clear
signatures of non-Gaussianity.} \label{whitened} 
\end{apjemufigure}

Using Eq.(\ref{pow}), one can easily visualize the 
$N_{\l-\frac{\Delta \l}{2},m}N_{\l+\frac{\Delta \l}{2},m}$
cros-correlations using simple definition
\begin{equation}
G_{\l m}(\Delta \l)=N_{\l-\frac{\Delta \l}{2},m}N^{*}_{\l+\frac{\Delta
    \l}{2},m}=\exp \left[iD_{\l m}(\Delta \l)\right]
\label{ccr}
\end{equation}
For these maps the common features, like a small in size clusters,
seems to be typical (corresponding statistics one can
find in \citet{ndv4}). In order to show how local defects of the maps
can produce corresponding features, let us introduce the following model of
sub-dominant non-Gaussian signal. Suppose that we have a set of points
(pixels) with coordinates $\theta_j,\phi_j$ in which the signal looks
like a combination of $\delta$-functions

\begin{equation}
\Delta T_{loc}(\theta,\phi) = \sum_j A_j \delta(\phi -\phi_j)
\delta(\cos\theta-\cos\theta_j),
\label{ps}
\end{equation}
where $A_j$ are the amplitudes of defects.
In order to obtain the corresponding $c_{\l m}$ coefficients, we
convolve $\Delta T$ from Eq.(\ref{ps}) with conjugated spherical harmonics
\begin{eqnarray}
c_{\l m}& = &\int_{-1}^{1}d(\cos\theta) \int_{-\pi}^{\pi}d\phi \Delta
T_{loc} (\theta,\phi)\Ylm^{*}(\theta,\phi)\nonumber \\
&=& \sum_i A_i \Ylm^{*}(\theta_i,\phi_i)
\label{psy}
\end{eqnarray}

\subsection{Peculiarities of $\Delta \l$-even statistics}
To obtain the correlators for $\Si$ and $\Cs$ statistics, we
need to know  $\cos[D_{\lm}(\Delta \l)]$ and $\sin[D_{\lm}(\Delta \l)]$, which
can be found from Eq.(\ref{psy}) as
\begin{eqnarray}
C_{\lm}(\Delta \l)&=&|c_{\l-\frac{\Delta \l}{2},m}||c_{\l+\frac{\Delta
    \l}{2},m} |\cos\left[D_{\lm}(\Delta \l)\right] \nonumber\\
&=&\left( c_{\l-\frac{\Delta \l}{2},m}c^{*}_{\l-\frac{\Delta \l}{2},m}+
c^{*}_{\l-\frac{\Delta \l}{2},m}c_{\l-\frac{\Delta \l}{2},m}\right)/2
    \nonumber\\ 
S_{\lm}(\Delta \l)&=&|c_{\l-\frac{\Delta \l}{2},m}||c_{\l+\frac{\Delta
    \l}{2},m} |\sin\left[D_{\lm}(\Delta \l)\right] \nonumber\\
&=&\left(c_{\l-\frac{\Delta \l}{2},m}c^{*}_{\l-\frac{\Delta \l}{2},m}-
c^{*}_{\l-\frac{\Delta \l}{2},m}c_{\l-\frac{\Delta \l}{2},m}
    \right)/2i \nonumber\\
\label{psyss}
\end{eqnarray}

Taking into account Eq.(\ref{psy}), from Eq.(\ref{psyss}) we have
\begin{eqnarray}
C_{\lm}(\Delta \l)&=&\sum_{j,k}
B_{jk}(\l,m)\cos\left[m(\phi_j-\phi_k)\right] \nonumber\\
&\times& P_{\l-\frac{\Delta \l}{2}}^m(\cos\theta_j)
P_{\l+\frac{\Delta \l}{2}}^m(\cos\theta_k) \nonumber\\
S_{\lm}(\Delta \l)&=&\sum_{j,k}B_{jk}(\l,m) \sin\left[m(\phi_j-\phi_k)
\right] \nonumber\\
&\times& P_{\l-\frac{\Delta \l}{2}}^m (\cos\theta_j)
P_{\l+\frac{\Delta \l}{2}}^m(\cos \theta_k)\nonumber\\
\label{CS}
\end{eqnarray} 
where $P_\l^m(\cos\theta)$ are the associated Legendre polynomials and 
\begin{eqnarray}
&&B_{jk}(\l,m)=\nonumber \\
&& A_j A_k\left[\frac{\Gamma(\l-\frac{\Delta \l}{2} - m+ 1)
    \Gamma(\l+ \frac{\Delta \l}{2} - m +1)}
{\Gamma(\l-\frac{\Delta \l}{2} +m+1)\Gamma(\l+\frac{\Delta \l}{2}
    +m+1)}\right]^{\frac{1}{2}} 
\label{CS1}
\end{eqnarray}

Following \citet{ndv3}, let us describe an ideal situation when
$\theta_j=\theta_k=\pi/2$, but $\phi_j\neq \phi_k$. 
For this model in Eq.(\ref{CS}) we get
\begin{eqnarray}
&&P_{l-\frac{\Delta \l}{2}}^m(0)= \nonumber \\
&& \sqrt{\pi}2^{m}
      \left[\Gamma\left(1+\frac{\l-m- \frac{\Delta
      \l}{2}}{2}\right) 
\Gamma\left(\frac{1-\l-m+\frac{\Delta \l}{2}}{2}\right) \right]^{-1} 
\label{leg}
\end{eqnarray}

As one can see from Eq.(\ref{leg}) and Eq.(\ref{CS}) the major part of
the summation over $m$ is related with $m=\l-\frac{\Delta \l}{2}$, 
for which 
\begin{eqnarray}
&&P_{\l-\frac{\Delta \l}{2}}^m(0) P_{\l+\frac{\Delta
    \l}{2}}^m(0)|_{m=\l-\frac{\Delta \l}{2}}=  \nonumber\\
&& 2^{2 \l-\Delta \l}\pi \frac{\Gamma\left(\frac{1}{2}+(\l-\frac{\Delta
    \l}{2}) \right) \Gamma(\frac{1}{2}+\l)}
    {\Gamma(\frac{1}{2}+\frac{\Delta \l}{2})} \cos(\frac{\pi\Delta \l}{2})
\label{leg1}
\end{eqnarray}


Thus, the sign of the $C_{\l m}(\Delta \l)$ for even  $\Delta \l$ 
is determined by the cosine terms. Moreover, if $\Delta \l=2n$,
$n=1, 2, 3\ldots$, the contribution of the local defects with
$\theta_j=\pi/2$ to $\Cs$ statistics manifest itself as the changes of
the sign, mentioned in Eq.(\ref{delt}). One may ask, why are $\Si$ and $\Cs$
statistics so different, if they depends on $\theta_j$ in the same
way? Taking into account the
$\phi_j$ dependency of $C_{\l m}(\Delta \l)$ and $S_{\l m}(\Delta \l)$
one can see that for $j=k$ the cosine terms in Eq.(\ref{CS})
($\cos \left[m(\phi_j-\phi_k)\right]$) goes to unity while the sine
terms for $S_{\l,m}(\Delta \l)$ is equal to zero.
All $j\neq k$ modes in Eq.(\ref{CS}) are represented by highly
oscillated functions and look like a noise in the phase diagrams
(see Fig.\ref{deltal2}-\ref{deltal4}).


\subsection{Peculiarities of odd $\Delta \l$ statistics}
Let us discuss the model $\Delta \l = 2n+1$, $n=1, 2, 3 \ldots$. There are
two possibilities to understand the properties of the phase correlations
for odd $\Delta \l$. Firstly we discuss the Galactic plane sources
contamination. In such a case similar to Eq.(\ref{leg1}) one can find
\begin{eqnarray} 
P_{\l-\Delta \l}^m(0)P_{\l}^m(0)|_{m=\l-\Delta \l}\sim \cos\left(\frac{\pi
  \Delta \l}{2}\right), 
\label{leg2}
\end{eqnarray}
and odd $\Delta \l$ does not contribute to $\Si$ and $\Cs$ statistics,
if $\theta_j=\pi/2$ for all $j$. However, if for some of the local
defects $\theta_j \neq \pi/2$, but $\cos \theta_j \ll 1$, there should produce
significant peculiarities in $\Si$ statistics by the following way. Close to the Galactic plane we can expand
the Legendre polynomials using a Taylor series
\begin{eqnarray}
P_{\l-\Delta \l}^m(\cos\theta_j)=P_{\l-\Delta
  \l}^m(0)+\frac{dP_{\l-\Delta \l}^m(cos \theta_j)}{d(\cos\theta_j)}
|_{\cos\theta_j=0} \cos\theta_j \nonumber\\
+ \frac{1}{2}\frac{d^2 P_{\l-\Delta \l}^m
  (\cos\theta_j)}{d(\cos\theta_j)^2}|_{\cos\theta_j=0} \cos^2 \theta_j
  \nonumber\\ 
\label{exp}
\end{eqnarray}
Simple algebra allows us to conclude that  the dependence $\S_{\lm}(\Delta
\l),\Cs_{\lm}(\Delta \l) \propto \cos^2\theta_j \sin(\pi \Delta
  \l/2)$  comes from interference between the first and the last
terms in Eq.(\ref{exp}), after their substitution to
Eq.(\ref{CS}). Thus, this dependency over $\Delta \l$
takesplace for some of the multipoles, for which $\l \theta_j \ll 1$,
while for $\l \theta_j \gg 1$
  $\S_{\lm}(\Delta \l),\Cs_{\lm}(\Delta \l)$ statistics goes to zero
  for odd $\Delta \l$. This dependency of the
$\S_{\lm}(\Delta \l),\Cs_{\lm}(\Delta \l)$ over $\delta \l$ can be clearly
seen in Fig.\ref{deltal1} and Fig.\ref{deltal3}, where symmetry $\Si$
and $\Cs$ statistics are restored for high $\l$. However,
$\S_{\lm}(\Delta \l)$ as $\Cs_{\lm}(\Delta  \l)$ depends on $\Delta
\l$ in the same way, while in Fig.\ref{deltal1} and Fig.\ref{deltal3}
one can see significant asymmetry in the $\Si$ and $\Cs$
statistics. The reason for that could be localization of the the
defects in $\phi$ direction close to the Galactic center with the
width $-\pi/2 \ll \delta\phi_j \ll \pi/2$ in the Galactic coordinates.

\section{Conclusion}
In this paper we examine the primordial non-Gaussian feature from
Alvan turbulence caused by primordial magnetic field. The distortion of the blackbody CMB spectrum by such homogeneous
magnetic field is not discussed in this paper. Due to the vector
nature of the magnetic field, off-diagonal correlations between
spherical harmonic modes $a_{\l-1,m}$ and $a_{\l+1,m}$ are induced. To
see the $\Delta \l =2$ correlations, we apply circular statistics of
phases to analyze such non-Gaussian component from the CMB signal. We
have applied the statistics on the ILC, FCM and WFCM maps, using all the
avaliable phases. The phase information can help us to test some
of the properties of the signal even when the amplitude of the power
spectra were smoothed by Gaussian window functions starting from $\l
\sim 50-200$. 

The FCM and WFCM maps have non-Gaussian signature registered in the
phases, as we have found phase correlations not only $\Delta \l=2$,
but also $\Delta \l=1$, 3 and 4. Such strong correlations are related
to the residues from component separation in the Galactic plane and
some point sources. None of the maps shows
specific features from the vortex perturbations
at the last scattering surface. Moreover, the detected non-Gaussianity
is completely different from the vortex perturbations properties. 
Roughly speaking, all the modes with $\Delta \l =1$, 2, 3 and 4  
show pronounced correlations, which is not the specific features
for the vortex. The correlations such as $\Delta \l=4, 8 \ldots$ might
have been the higher order correlation from $\Delta \l=2$, but the
$\Delta \l=1$  and others indicate that such strong correlations are
related to the residues from foreground component separation. We
conclude that the Alfven turbulence does not
contribute significantly in the phase information. The method we
propose for phase analysis is useful for the upcoming PLANCK mission,
especially for testing the properties of foreground cleaning
methods. Our final remark is that the MHD structure claimed by
\citet{ber03} and \citet{bs04} in the WFCM map is related with
Galactic plane residues but not with the Alfven turbulence contamination. 


\acknowledgements
We thank Tegmark et al. for providing their processed maps. 
We acknowledge the use of \healpix \footnote{\tt
 http://www.eso.org/science/healpix/} package \citep{healpix} to
produce $\alm$ from the \wmap data and the use of the \glesp package
\citep{glesp} for data analyses and the whole-sky figures.


\end{document}